\documentclass[11pt]{article}

\usepackage{epsfig}
\usepackage{amssymb}
\usepackage{natbib}
\usepackage[hyperfootnotes=false]{hyperref}
\usepackage{graphicx}
\usepackage{thumbpdf}
\usepackage{amsfonts,amstext,amsmath}
\usepackage{accents}
\usepackage{color}
\usepackage{rotating}
\usepackage[utf8]{inputenc}
\usepackage{booktabs}
\usepackage{tikz}
\usepackage{multirow}
\usepackage{float}
\usepackage[labelfont=bf, width=15cm, font=small]{caption}
\usepackage[onehalfspacing]{setspace}
\usepackage[left=3cm, right=3cm, top=2.5cm, bottom=3cm]{geometry}
\setcitestyle{authoryear,open={(},close={)}}
\newcommand*{\doi}[1]{\href{http://dx.doi.org/#1}{doi: #1}}

\usepackage{accents}

\newcommand{\rZ}{Z}
\newcommand{\rY}{Y}
\newcommand{\rX}{\mX}

\newcommand{\rz}{z}
\newcommand{\ry}{y}
\newcommand{\rx}{\xvec}

\newcommand{\pZ}{F_\rZ}

\newcommand{\pYx}{F_{\rY | \rX = \rx}}

\newcommand{\dZ}{f_\rZ}

\newcommand{\dYx}{f_{\rY | \rX = \rx}}

\newcommand{\h}{h}

\newcommand{\basisy}{\avec}
\newcommand{\bern}[1]{\avec_{\text{Bs},#1}}

\newcommand{\shiftparm}{\betavec}

\newcommand{\eshiftparm}{\beta}

\newcommand{\ie}{\textit{i.e.,}~}
\newcommand{\eg}{\textit{e.g.,}~}

\newcommand{\Prob}{\mathbb{P}}

\usepackage{dsfont}

\newcommand{\given}{\rvert}

 \DeclareMathOperator{\expit}{expit}

\def \avec {\text{\boldmath$a$}}

\def \xvec {\text{\boldmath$x$}}

\def \rX {\text{\boldmath$X$}}

\def \betavec         {\text{\boldmath$\beta$}}

\def \thetavec        {\text{\boldmath$\theta$}}

\newcommand{\ubar}[1]{\underaccent{\bar}{#1}}

\newcommand{\proglang}[1]{\textsf{#1}}
\newcommand{\code}[1]{\texttt{#1}}

\def \B {\mathsf{B}}
\def \linpred {\rx^\top\shiftparm}

\newcommand{\ls}{LS}
\newcommand{\cs}{CS}
\newcommand{\si}{SI}
\newcommand{\ci}{CI}
\newcommand{\lsx}{\ls$_\rx$}
\newcommand{\lsmRS}{\ls$_{\rm mRS}$}

\newcommand{\cib}{\ci$_\B$}
\newcommand{\csb}{\cs$_\B$}
\newcommand{\silsx}{\si-\lsx}

\newcommand{\sicsblsx}{\si-\csb-\lsx}
\newcommand{\ciblsx}{\cib-\lsx}
\usepackage{tikz}

\usetikzlibrary{positioning, calc, shapes.geometric, shapes.multipart,
  shapes, arrows.meta, arrows,
  decorations.markings, external, trees}
\tikzstyle{line} = [draw, -latex']
\tikzstyle{Arrow} = [
        thick,
        decoration={
                markings,
                mark=at position 1 with {
                        \arrow[thick]{latex}
} },
        shorten >= 3pt, preaction = {decorate}
        ]

\newcommand\pkg[1]{\textbf{#1}}

\newcommand\revision[1]{{\color{black}#1}}

\begin{document}

\title{\bf Deep transformation models for functional outcome prediction after acute ischemic stroke}
\author{%
    Lisa Herzog\textsuperscript{1,2,3,}{\small *},
    Lucas Kook\textsuperscript{1,2,}{\small *},
    Andrea G\"otschi\textsuperscript{1},
    Katrin Petermann\textsuperscript{1},
    \\
    Martin H\"ansel\textsuperscript{3},
    Janne Hamann\textsuperscript{3},
    Oliver D\"urr\textsuperscript{4},
    Susanne Wegener\textsuperscript{3},
    Beate Sick\textsuperscript{1,2}%
}
\date{
   \footnotesize%
   \textsuperscript{1}Epidemiology, Biostatistics \& Prevention Institute, University of Zurich, Switzerland\\
   \textsuperscript{2}Institute for Data Analysis and Process Design, Zurich University of Applied Sciences, Switzerland\\
   \textsuperscript{3}Department of Neurology, University Hospital Zurich, Switzerland\\%
   \textsuperscript{4}Insitute for Optical Systems, Konstanz University of Applied Sciences%
}

\maketitle

\begin{abstract}%
In many medical applications, interpretable models with high prediction performance are sought. Often, those models are required to handle semi-structured data like tabular and image data. We show how to apply deep transformation models (DTMs) for distributional regression which fulfill these requirements. DTMs allow the data analyst to specify (deep) neural networks for different input modalities making them applicable to various research questions. Like statistical models, DTMs can provide interpretable effect estimates while achieving the state-of-the-art prediction performance of deep neural networks. In addition, the construction of ensembles of DTMs that retain model structure and interpretability allows quantifying epistemic and aleatoric uncertainty. In this study, we compare several DTMs, including baseline-adjusted models, trained on a semi-structured data set of 407 stroke patients with the aim to predict ordinal functional outcome three months after stroke. We follow statistical principles of model-building to achieve an adequate trade-off between interpretability and flexibility while assessing the relative importance of the involved data modalities. We evaluate the models for an ordinal and dichotomized version of the outcome as used in clinical practice. We show that both, tabular clinical and brain imaging data, are useful for functional outcome prediction, while models based on tabular data only outperform those based on imaging data only. There is no substantial evidence for improved prediction when combining both data modalities. Overall, we highlight that DTMs provide a powerful, interpretable approach to analyzing semi-structured data and that they have the potential to support clinical decision making. 
\footnote{{\small *}Authors contributed equally \newline
Preprint; under review. Version: September 13, 2022. Licensed under CC-BY.}
\end{abstract}

\clearpage

\section{Introduction} \label{sec:intro}

\revision{%
Although prevention, diagnosis and treatment of stroke have improved largely, 
it remains one of the leading causes of long term disability and death worldwide 
\citep{AhaUpdate2019}. Each year, approximately 15 million people experience a stroke, 40\% 
die and 30\% suffer lasting functional disability. To achieve the best possible 
outcome, patients have to be treated as fast as possible and decisions 
for or against different treatment options have to be made under immense time pressure.
For clinical studies, the patient's functional outcome three months after hospital admission
is primarily used to assess treatment success. Functional outcome is quantified on the
modified Rankin Scale (mRS), an ordinal score comprising seven levels ranging between no
symptoms at all (mRS of 0) and death \citep[mRS of 6,][]{quinn2009reliability}. 
Often, neurologists are not directly interested in predicting the exact class of the 
mRS but rather in stratifying the chances of a patient having a favorable (mRS of 0--2) vs. 
unfavorable (mRS of 3--6) functional outcome \citep{weisscher2008should}.}

Semi-structured data comprise the basis for various decisions in medicine 
\citep[\eg in stroke and cancer,][]{ebisu1997hemorrhagic,jafari2018breast}. For 
instance, when predicting functional outcome in stroke patients, unstructured data 
such as brain images resulting from Computed Tomography (CT) or Magnetic Resonance 
Imaging (MRI) are as important as structured data, like tabular patient and clinical
characteristics \citep{Copen2011}. 
Different brain imaging modalities provide insight into the extent of tissue injury, 
the exact location of the stroke lesion as well as previous brain infarcts. 
While in clinical practice, information from brain imaging is frequently 
used for difficult clinical decisions, functional outcome prediction is limited with 
current image analysis strategies (see Section~\ref{sec:relatedwork}). 
\revision{%
It is currently an open question to what extent the imaging data and tabular data help
in reliably predicting functional outcome. In a previous study, \citet{HamannHerzog2021}
found no additional benefit for stroke outcome prediction when adding expert-derived image
features alongside clinical features.
Trustworthy models for outcome prediction relying on data of both modalities
are lacking but of high interest to the neurologist to assess the
vast amount of complex medical data under immense time pressure.
} 

Recently, machine learning (ML) and deep learning (DL) models in particular, 
have proven outstanding prediction results on unstructured data like images. The models 
are fast, precise and reproducible when it comes to analyzing the large amount of 
data appearing in daily clinical practice \citep[\eg][]{Campanella2019}. 
Nonetheless, there is often distrust in ML derived predictions, which is mainly due to 
their ``black-box'' character \citep{rudin2018stop}. Questions like ``How does the model 
come to its prediction?'', ``How certain is the model about the prediction?'', or ``What 
is the impact of different patient features on the prediction'' have to be answered, in 
order for medical experts to trust the model. Therefore, ML models should not only focus 
on achieving the most accurate predictions but also on interpretability and uncertainty, 
\ie the models should be tailored to provide a distributional outcome prediction instead 
of a point prediction.

We present deep transformation models (DTMs) to analyze semi-structured data. DTMs unite 
classical statistical models with (deep) neural networks, provide distributional outcome 
predictions, and achieve interpretable model parameters without sacrificing the high 
prediction performance of deep learning models. We demonstrate the use of DTMs on data of 
patients admitted to the hospital due to stroke symptoms. In particular, we present 
models for predicting a patient's functional outcome measured by the ordinal mRS 
three months after hospital admission that rely on tabular data, brain imaging or 
a combination of both. 
\revision{%
We apply DTMs on a semi-structured data set of 407 stroke patients to model the 
conditional distribution of a patient's functional outcome three months after hospital
admission. We describe briefly how DTMs can be used to model continuous or censored 
outcomes, like time-to-event data, which makes them applicable to many different research 
questions. We discuss how DTMs yield interpretable effect estimates of the different input
modalities and how the model arrives at its predictions. Moreover, we highlight 
baseline-adjusted DTMs conditioning on a patient's pre-stroke mRS, which is expected to
be strongly predictive of outcome. Baseline-adjusted DTMs for un- and semi-structured data
are novel and of high interest to data analysts working in medical research, in which
integrating baseline variables for outcome prediction is a common requirement.
}

\subsection{Related work} \label{sec:relatedwork}

In the following, we describe work related to semi-structured distributional regression 
approaches as DTMs.

\paragraph{Classical regression models} Classical regression models like logistic 
regression or Cox proportional hazard models are the standard when analysing structured 
data (\eg tabular features) in medical applications. They are considered highly trustworthy because 
they are transparent, interpretable and provide uncertainty measures \citep[\eg][]{steyerberg2019}. 
However, unstructured data like images or text cannot directly be analyzed with such models. 
First, tabular features have to be extracted from the unstructured data to be subsequently analyzed
in a regression model -- potentially together with other tabular data \citep[\eg][]{thiran1996feature}.
Yet, this features engineering step is disconnected from optimizing the model parameters and necessarily 
discards information, which makes it difficult to know if the engineered features reflect relevant 
information in the original data well enough.

\paragraph{Deep neural networks} (Deep) neural networks (DNNs), on the other hand, 
learn relevant features for a task at hand as a part of the model fitting process and 
therefore omit the feature engineering step while they can be trained 
on structured data, unstructured data or a combination of both \citep{goodfellow2016deep}. For instance, 
previous work has focused on analyzing combinations of image and tabular data to predict 
stroke patient outcomes with DNNs. \citet{Pinto2018} used a 
model consisting of a convolutional neural network (CNN) for the image data where they 
attach tabular data to the feature vector in the dense part of the CNN. This enables 
interactions between image and tabular data. Another pilot study for stroke outcome 
prediction used a combination of a CNN and a dense NN for integrating image and tabular 
data into one model \citep{Bacchi2020}. 
However, like the majority of the DNNs, the existing approaches are black box models which 
do not quantify uncertainty. They lack interpretable model parameters and estimate point 
predictions like the conditional mean rather than a conditional outcome distribution.

\paragraph{Distributional regression} Distributional regression focuses on 
estimating an entire conditional distribution rather than the first conditional 
moment(s) \citep{kneib2021rage}. \revision{Therefore, when fitted by empirically optimizing
a proper score like the negative log likelihood, a distributional regression model directly 
quantifies aleatoric uncertainty inherent in the data}. 
To achieve a well fitting distributional regression model, a complex conditional outcome 
distribution might be required. Generalized linear models (GLM) are based on members of 
the exponential family, defined by the first two moments, for modeling the conditional 
outcome distribution while they provide interpretable model parameters.
Generalized Additive Models for Location, Scale and Shape (GAMLSS) extend GLMs by 
allowing to specify all parameters of the assumed outcome \citep{stasinopoulos2007gamlss}.
A GAMLSS implementation with flexible specification of the conditional moments of 
$\Prob_{\rY\given\rX=\rx}$ using deep neural networks is, for example, presented in
\citep{rugamer2020unifying}. However, these models still require the choice of 
a parametric family of conditional outcome distributions.

\paragraph{Transformation models for distributional regression} Transformation
models (TMs) are a more recent method for distributional regression, which do
not require to pre-specify the family of the outcome distribution
\citep{hothorn2014conditional,hothorn2018most}. In TMs the conditional
outcome distribution is decomposed into a simple, parameter free,
target distribution $\pZ$ (\eg normal or logistic) and a conditional transformation 
function $\h(\ry \given \rx)$, such that $\pYx(\ry) = \pZ(\h(\ry\given\rx))$.
More details are given in Section~\ref{sec:methods}. Independent of TMs, normalizing 
flows were developed in the deep learning community \citep{rezende2015}, which are 
based on the same idea as TMs. But while normalizing flows solely aim at predicting a 
flexible (conditional) distribution and constructing the transformation function as a 
chain of simple transformations, TMs are tailored for interpretable distributional 
regression models. The construction of the transformation function and the choice of 
the simple distribution $\pZ$ give rise to extremely flexible TMs for conditional 
distributions. For instance, \citet{sick2020deep} and \citet{baumann2020deep} use 
$\pZ = \Phi$ and predict different outcome distributions with variously flexible 
transformation functions on commonly used benchmark data sets in deep learning and 
demonstrate state-of-the-art prediction performances. \citet{rugamer2021timeseries} 
use DTMs for time series data by including auto-regressive components in the 
transformation function. 

\paragraph{(Deep) transformation models for ordinal outcomes} The main 
application of this article features an ordinal outcome (mRS). Models for the conditional 
distribution of an ordinal outcome given covariates like the proportional odds logistic 
regression model have been studies in statistics for several decades 
\citep{mccullagh1980regression}. Baseline-adjusted proportional odds models 
have been described from a transformation-model perspective in \citet{Buri2020}. However,
only recently a special DTM, focusing on ordinal neural network transformation models 
({\sc ontram}s) has been developed in deep learning and applied to several publicly
available (non-medical) data sets (Kook \& Herzog et al., \citeyear{kook2020ordinal}). 
However, DTMs were not yet applied in the context of stroke. 

\revision{%
\paragraph{Transformation ensembles}
Ensembling in terms of aggregating the predictions of multiple models to improve prediction
performance is commonly seen in practical applications. In the field of deep learning,
ensembling often means aggregating the predicted probabilities of a few DNNs that possess
the same architecture and are trained on the same data after random initialization
\citet{lakshminarayanan2016deepensembles}. These deep ensembles are not only used to
achieve more accurate predictions but also to quantify epistemic uncertainty by means
of the variation of the different predictions. However, the special structure and the
interpretability of deep TMs are in general lost after aggregating them via deep ensembling.
\citet{kook2022interpretable} recently developed transformation ensembles which aggregate DTMs
on the scale of the transformation function preserving structure and interpretability
(see Section~\ref{sec:methods}).
}

This article is organized as follows. Section~\ref{sec:methods} presents detailed background
on distributional regression models with semi-structured data and the experimental setup
including model evaluation. Results are presented in Section~\ref{sec:results}. We end with
a discussion of the various types of questions that may be answered by deep distributional
regression models like DTMs in Section~\ref{sec:outlook}.

\section{Methods} \label{sec:methods}

In the following, we briefly introduce TMs which are used to integrate semi-structured data,  
model highly flexible conditional outcome distributions, and provide 
interpretable model parameters. Since our application
features an ordinal outcome, we will pay special attention to this case.

\subsection{Distributional regression with transformation models} \label{sec:dctm}

\begin{figure}[!ht]
\center
\resizebox{\textwidth}{!}{%
\begin{tikzpicture}
\node (a) {\includegraphics[width=0.45\textwidth]{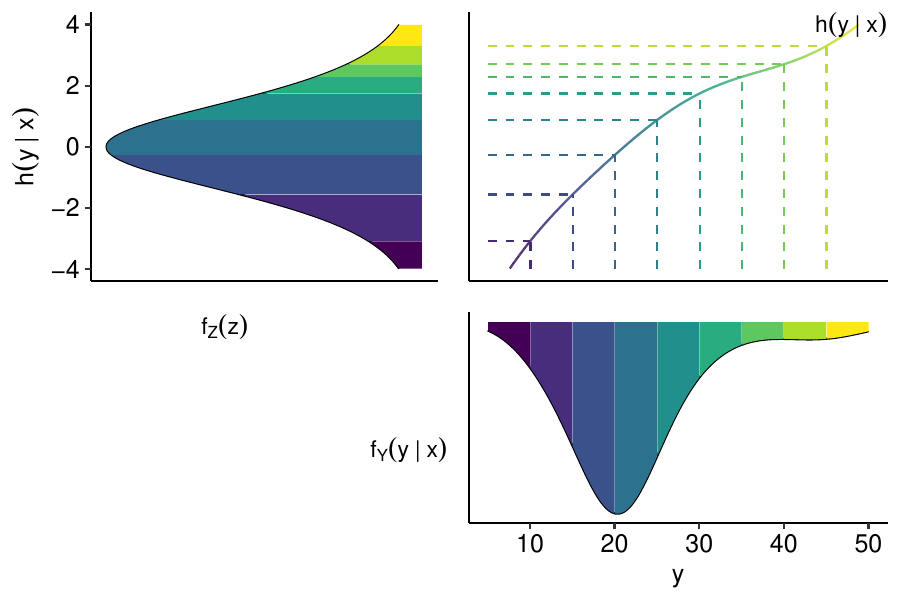}};
\node[right=0.5cm of a] (b) {\includegraphics[width=0.45\textwidth]{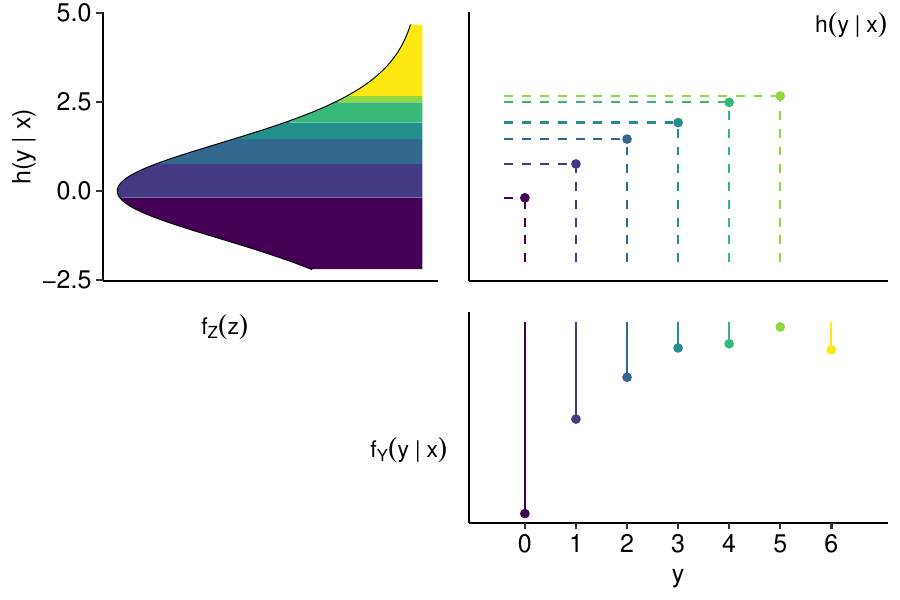}};
\node[above left=0cm of a] (taga) {\textsf{a}};
\node[above left=0cm of b] (tagb) {\textsf{b}};
\end{tikzpicture}}
\caption{
  TMs for continuous (\textsf{a}) or ordinal categorical outcome (\textsf{b}).
  The lower right part of each panel 
  shows the conditional density of $\rY$ given $\rx$, which is mapped onto the
  density of the latent variable $\rZ$ (see upper left part in each panel). The
  transformation is done via a monotone transformation function $\h$ (upper
  right part). This transformation function, can be continuous (\textsf{a}) or
  discrete (\textsf{b}). 
}\label{fig:trafo}
\end{figure}

In TMs the problem of estimating the potentially complex conditional outcome distribution
of $(\rY \given \rX = \rx)$ is approached by learning a parameterized monotone
transformation $\h(\ry \given \rx; \thetavec)$ which maps between the distributions of 
$(\rY \given \rX = \rx)$ and the latent variable $\rZ$. The distribution of the latent
variable $\rZ$ (with log-concave density) has to be defined \emph{a priori}.
Usually, a parameter-free distribution, such as the standard Gaussian or
logistic distribution, is chosen (see Fig.~\ref{fig:trafo}). 

The parameters $\thetavec$ in $\h(\ry \given \rx; \thetavec)$ determine the
functional form of the transformation function and thus the corresponding
conditional outcome distribution (we drop $\thetavec$ in the following to
simplify notation). The parameters are fitted via maximum likelihood, \ie
by minimizing
\begin{align}\label{eq:nll}
\text{NLL} = - \frac{1}{n} \sum_{i=1}^n \ell_i(\thetavec),
\end{align}
where $\ell_i$ is the log-likelihood contribution of the $i$-th training
observation.

In case of a continuous outcome, the likelihood contribution of an exact
observation $(\ry, \rx)$ is given by the value of the conditional density at
the observed outcome $\dYx(\ry)$ which can be determined via $\dZ$ and $\h$
by using the change of variables formula $\dYx(\ry) = \dZ(\h(\ry\given \rx))
\cdot \h'(\ry\given \rx)$. The transformation function $\h$ is a smooth
function (see Fig.~\ref{fig:trafo}\textsf{a}) which can be modeled via a basis
expansion with basis functions $\avec(\cdot)$, yielding $\avec(\ry)^\top\thetavec$.
A common choice for $\avec(\cdot)$ are polynomials in Bernstein form $\bern{P}(\ry)$ 
of order $P$. Here the required monotonicity of $\h$ can be easily guaranteed via linear
constraints on the parameters $\thetavec$ \citep{hothorn2018most}.
Complex dependence on the input $\rx$ can be achieved by controlling 
$\thetavec(\rx)$ via a deep NN. 

If a continuous observation is censored, which occurs especially often in
survival data, the outcome is measured as an
interval $\ry \in (\ubar{\ry}, \bar{\ry}]$ and the likelihood contribution
can be derived from the cumulative distribution function, as
$\pZ(\h(\bar{\ry}\given \rx)) - \pZ(\h(\ubar{\ry}\given \rx))$.

For an ordered categorical outcome, the discrete monotone increasing
transformation function $\h$ maps the observed outcome classes
$(\ry_k, \rx)$ to the conditional cut points $\h(\ry_k \given \rx), \; k =
1, \dots, K - 1$ of the latent variable $\rZ$, as illustrated in
Fig.~\ref{fig:trafo}\textsf{b}. This allows to view the ordinal outcome as
result of an underlying continuous latent variable $\rZ$ with interval-censored
observations. The likelihood contribution of an observation
$(\ry_{k},\rx)$, given by the probability $p_k$ for the observed class
$\ry_{k}$, can correspondingly be determined by the area under $\dZ$ between the
cut points $\h(\ry_{k} \given \rx)$ and $\h(\ry_{k-1} \given \rx)$ and is
computed as
$p_k = \pZ(\h(\ry_{k} \given \rx)) - \pZ(\h(\ry_{k-1} \given \rx))$. If
dummy-encoding is used for $\ry_k$, \ie the class $k$ is encoded by a vector
$\basisy(\ry)$ of length $K$ which holds a \emph{one} at position $k$ and
\emph{zeros} elsewhere, then $\h$ is given by $\h(\ry_k \given
\rx)=\basisy(\ry)^\top \thetavec(\rx)$ with $\thetavec(\rx)$ being constrained
to $\theta_1(\rx) \leq \theta_2(\rx) \leq \dots \leq
\theta_K(\rx)=+\infty$.

\subsubsection{Interpretability in transformation models}
\label{sec:interpretation}

To achieve the same interpretability as in commonly used regression models, such
as proportional hazard or proportional odds models, the flexibility of $\h$
needs to be restricted. This can be done by decomposing $h$ in a baseline transformation 
(intercept function) $\h_0$ which does not depend on the input data and one or several
shift terms $\h(\ry\given \rx) = \h_0(\ry) - \text{shift}(\rx)$. In such a shift
model, $\h_0$ determines the shape of the transformation function
$\h$ and only the shift terms depend on $\rx$, moving $\h$ up and down (see Fig.~\ref{fig:trafo}). 
A particularly simple example is a linear shift model of some tabular input data 
$x_j, \; j \in 1, \dots , J$, which looks as follows for a continuous outcome
$\h(\ry \given \rx) = \h_0(\ry) - \linpred$. Depending on the chosen
distribution for $\rZ$ the parameters $\shiftparm$ have a straightforward
interpretation. A summary of commonly used distributions for $\pZ$ and the
corresponding interpretational scales is given in \citet{siegfried2020count}.

When choosing \eg the minimum extreme value distribution for $\rZ$,
\ie $\pZ(\rz) = 1 - \exp(-\exp(\rz))$, the parameters $\eshiftparm_j, j =
1,\dots, J$ can be interpreted as log hazard-ratios. A well-known example is the
proportional hazard model that is often used for survival analysis, where the
bounded continuous outcome is a survival time. Survival analysis poses
additional challenges. For instance, usually not all patients experience the
event of interest during follow-up, leading to (right-) censoring with $\ry
\in (\ubar{y},+\infty)$, which can be easily handled in TMs, as described
above.

\begin{figure}[!ht]
    \center
    \begin{tikzpicture}[auto, Arr/.style={-{Latex[length=2.5mm]}}]
    \node (int1) {$\theta_1$};
    \node[below=0cm of int1] (int2) {$\theta_2$};
    \node[below=0cm of int2] (int3) {$\vdots$};
    \node[below=0cm of int3] (intK) {$\theta_M$};
    \node[left=1cm of int2, circle, draw] (1) {1};
    \draw[Arr] (1.east) -- (int1);
    \draw[Arr] (1.east) -- (int2);
    \draw[Arr] (1.east) -- (intK);
    \draw ($(1.north west)+(-0.5,0.9)$) rectangle ($(intK.south east)+(0.5,-0.25)$);
    \node[below right=1cm of intK.south west] (trafo) 
    	{$\h(\ry \given \B, \rx) \; = \; \basisy(\ry)^\top\thetavec \; -  \; \eta(\B) - \; \rx^\top\shiftparm \;$};
    \node[below=1cm of trafo] (cnn) {\includegraphics[height=2cm]{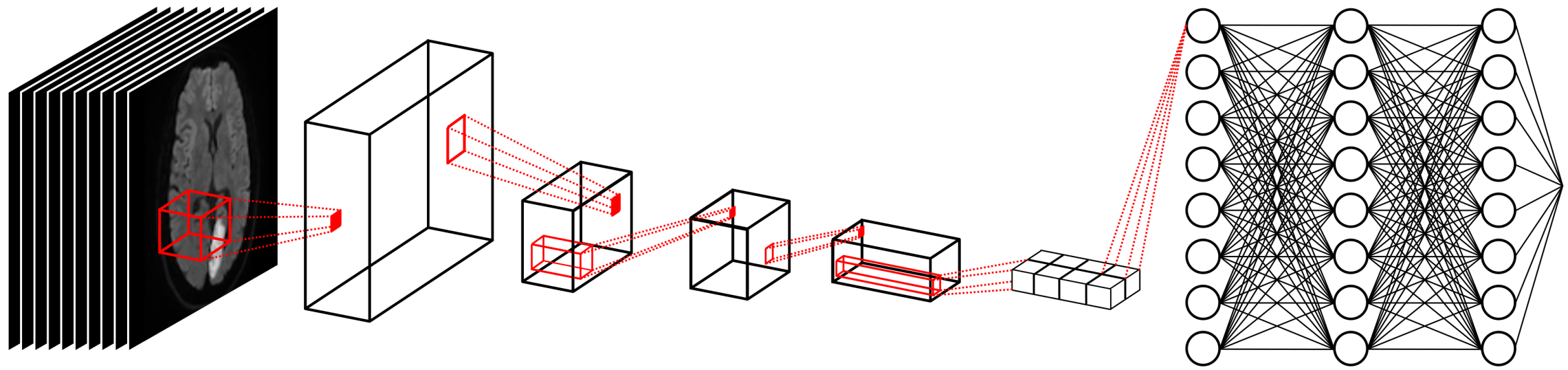}};
    \node[left=0cm of cnn] (B) {$\B$};
    \node[right=-0.1cm of cnn] (eta) {$\eta(\B)$};
    \draw ($(cnn.north west)+(-0.5,0.25)$) rectangle ($(eta.south east)+(0.5,-0.9)$);
    \node[right=6cm of int1] (x1) {$x_1$};
    \node[right=6cm of int2] (x2) {$x_2$};
    \node[right=6.5cm of int3] (x3) {$\vdots$};
    \node[right=6cm of intK] (xp) {$x_p$};
    \node[right=10 of 1] (linpred) {$\rx^\top\shiftparm$};
    \draw[Arr] (x1) -- node [midway, above] {\small $\beta_1$} (linpred.north west);
    \draw[Arr] (x2) -- node [midway, above] {\small $\beta_2$} (linpred.west);
    \draw[Arr] (xp) -- node [midway, above] {\small $\beta_p$} (linpred.south west);
    \draw ($(x1.north west)+(-0.5,0.25)$) rectangle ($(linpred.south east)+(0.5,-1.25)$);
    \draw[Arr] ($(intK.south east)+(0.5,-0.25)$) -- ($(trafo.north)+(-0.3,0)$);
    \draw[Arr] ($(cnn.north)+(0,0.25)$) -- ($(trafo.south)+(0.8,0)$);
    \draw[Arr] ($(xp.south)+(-0,0.03)$) -- ($(trafo.north east)+(-0.5,-0.00)$);
    \end{tikzpicture}
    \caption{A SI-CS$_\B$-LS$_\rx$ DTM with simple intercepts, not depending on the input,
    a linear shift term in the tabular input data $\rx$  and a 
    complex shift term in the images $\B$. All additive components of $\h$ are controlled by NNs, 
    \ie shallow dense NNs without hidden layer for SI and LS$_\rx$ and a three-dimensional 
    CNN for CS$_\B$, which are jointly fitted by minimizing the NLL via stochastic gradient
    decent.} 
    \label{fig:dctm}
\end{figure}
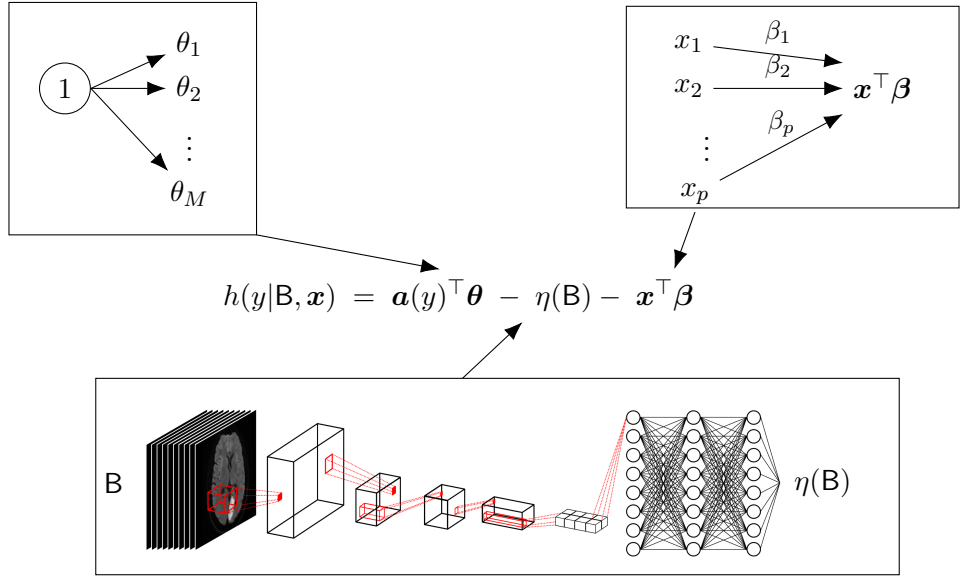

\paragraph{Semi-structured regression} In semi-structured regression, the
problem is to combine both structured data, \eg tabular features $\rx$, and
unstructured data, \eg images $\B$, in one single model. This can be
realized with NNs, which take both structured and unstructured data as
input and control the parameters of $\h$ (see Fig.~\ref{fig:dctm}).
Depending on the architecture of the NNs, more or less flexible models can be
described. 

The most flexible model is achieved, if $\h$ depends in complex manner on all
inputs corresponding to a complex intercept model with 
\begin{align}
\h(\ry \given \B, \rx) = \basisy(\ry)^\top\thetavec(\B, \rx), 
\end{align} 
where a NN controls $\thetavec(\cdot)$ depending on imaging data ($\B$) and tabular
data ($\rx$) and thus potentially allowing for interactions between $\B$ and $\rx$.
Without restricting $\thetavec(\cdot)$ in any way besides being monotone increasing,
maximal flexibility is achieved. Most often in biostatistics, a shift model is assumed
for $\h$ (\ie a proportionality assumption is made) and no interactions between the
input data are allowed. In this scenario, the model simplifies to 
\begin{align}
\h(\ry \given \B, \rx) = \basisy(\ry)^\top\thetavec - \eta(\B) - \beta(\rx),
\end{align} 
where $\beta$ and $\eta$ are controlled by two separate NNs and are interpretable, 
\eg as log odds-ratios if the logistic
distribution $F_Z(z)=\expit(z)=\frac{1}{1+\exp(-z)}$ is chosen. If a
linear effect is assumed for each tabular feature, and the effect of each
feature should be interpretable as log odds-ratio, then further
simplifications have to be made by using a linear shift term for the tabular data
\begin{align} 
    \h(\ry \given \B, \rx) = \basisy(\ry)^\top\thetavec - \eta(\B) -
    \rx^\top\shiftparm. 
\end{align} 
Such a model with simple intercept
$\basisy(\ry)^\top\thetavec$, linear shift $\rx^\top\shiftparm$, and complex shift
$\eta(\B)$ term is depicted in Fig.~\ref{fig:dctm} and referred to as
SI-CS$_\B$-LS$_\rx$ in this work.

In general, the primary goal is to develop a model with adequate prediction performance. 
Usually, simpler (\ie fewer parameter) and more interpretable models are preferred over 
black boxes. Only if the more complex model yields a substantial improvement in terms of 
prediction performance, the more complex model should be preferred. We investigate the
ramifications of model selection in Section~\ref{sec:results}.

\revision{%
\paragraph{Transformation ensembles}
We construct transformation ensembles of DTMs which are fitted on the same data but with
different random initialization. Transformation ensembles average the predicted transformation
functions of the DTMs, which preserves the model structure and interpretability, improves
prediction performance, and allows to quantify epistemic uncertainty \citep{kook2022interpretable}.
}
 
\subsection{Data} \label{sec:data}

\begin{figure}[!ht]
    \centering
    \includegraphics[width=0.7\textwidth]{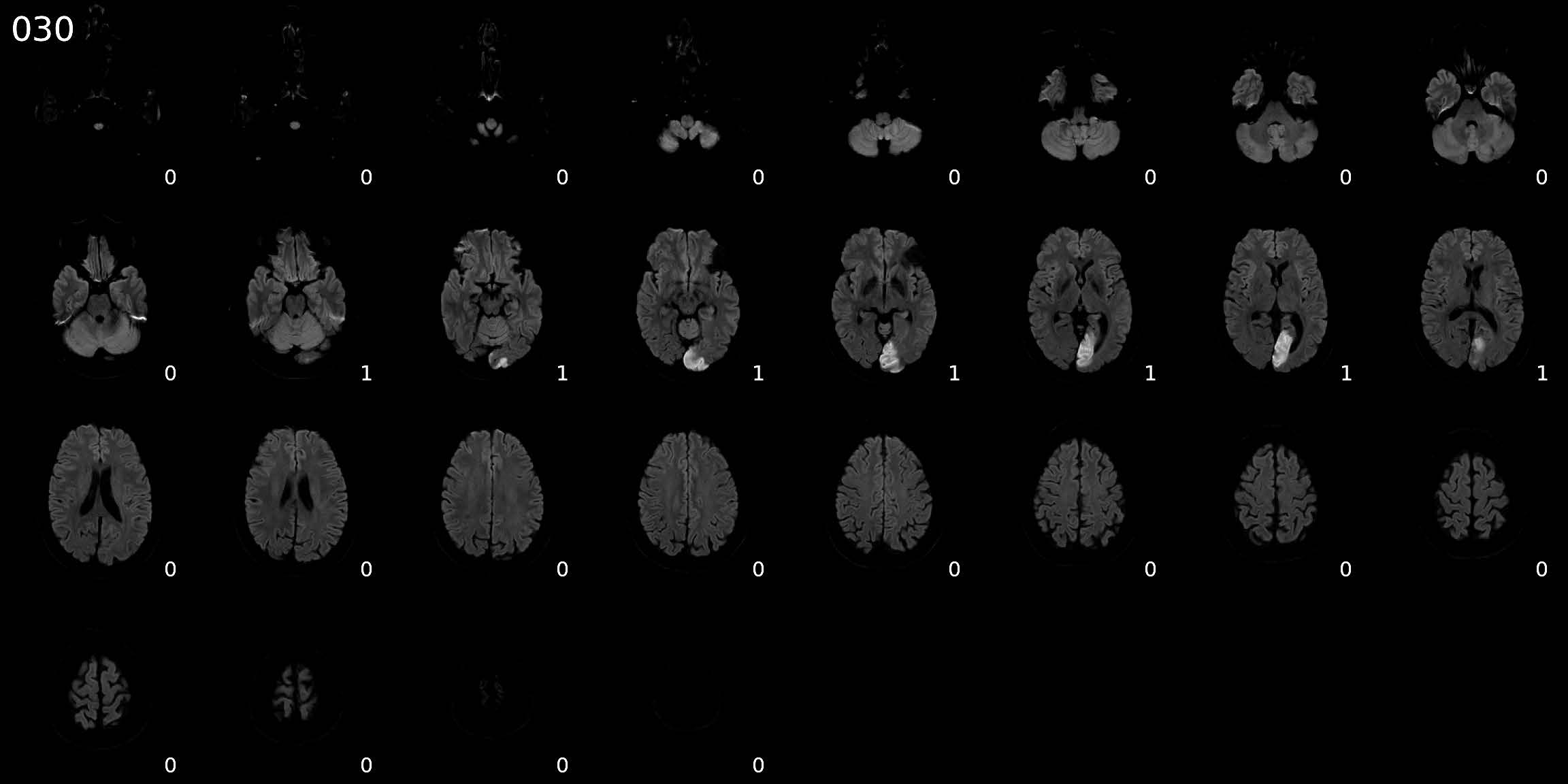}
    \caption{A pseudo 3D diffusion weighted image of an example stroke patient. 
    2D slices where a stroke lesion is visible are labeled with a 1 and 0 otherwise. 
    Each patient is represented by 28 diffusion weighted images (DWIs) after 
    pre-processing. Ischemic stroke lesions appear as hyper-intense signal on one or 
    multiple images of a sequence.}
    \label{fig:exampledwis}
\end{figure}

\revision{%
Our cohort consists of 407 patients who are either diagnosed with ischemic stroke 
(295 patients) or transient ischemic attack (TIA, 112 patients). As opposed to stroke, TIA
causes only temporary stroke symptoms and no permanent brain damage. The cohort was collected 
retrospectively. All patients were admitted to the University Hospital of Zurich between 2014 
and 2018 and had MRI records in the acute phase. Ethical approval for the study was obtained
from the Cantonal Ethics Committee Zurich (KEK-ZH-No. 2014-0304). 

In this study, we use the stroke patient's brain imaging and tabular baseline data for
functional outcome prediction. Diffusion Weighted Images (DWIs) represent brain pathology
in a 3D manner as ordered sequences of multiple 2D images per patient. On DWIs,
stroke lesions appear as hyper-intense signals, typically on multiple, subsequent images
in the sequence (see Fig.~\ref{fig:exampledwis}). They give valuable insight into stroke 
location and severity. TIA patients show no visible lesion on DWIs. All collected DWIs 
were recorded within three days after hospital admission.
After preprocessing, each 3D image is of dimension $128 \times 128 \times 28$ 
with zero mean and unit variance (see Fig.~\ref{fig:exampledwis}).
We consider baseline covariates, \ie patient characteristics including age and sex, risk 
factors including hypertension, prior stroke, smoking, atrial fibrillation, coronary heart 
disease (CHD), prior transient ischemic attack (TIA), diabetes and hypercholesterolemia, 
the National Institutes of Health Stroke Scale at baseline (NIHSS at BL) 
highlighting stroke symptom severity as an ordinal sum score with 42 levels, and the 
pre-stroke mRS (mRS at BL) informing about the patient's functional disability 
before stroke. All factor variables are dummy encoded and all other tabular features are 
standardized to make the magnitude of estimated parameters comparable.

The outcome of interest is the ordinal mRS, which consists of seven levels: 0 = no symptoms 
at all, 1 = no significant disability despite symptoms, 2 = slight disability, 
3 = moderate disability, 4 = moderately severe disability, 5 = severe disability,
6 = death \citep{grotta2016stroke}. 
In our cohort of 407 patients we observed the following $n_k, k = 1, \dots, K$ for
the $K = 7$ outcome classes: $n_0 = 184$ (45.2\%), $n_1 = 88$ (21.6\%), $n_2 = 60$ (14.7\%),
$n_3 = 25$ (6.1\%), $n_4 = 20$ (4.9\%), $n_5 = 5$ (1.2\%), $n_6 = 25$ (6.1\%).
Fig.~\ref{fig:predictors} in the Appendix shows
the distribution of predictors stratified by the outcome among all 407 patients.
Since in clinical practice, the neurologists are often primarily interested in the patient's
chance for a favourable (mRS $\leq 2$, $n_f = 332$, 81.6\%) vs. unfavourable 
(mRS $> 2$, $n_u = 75$, 14.6\%) outcome \citep{weisscher2008should}, we additionally 
considered the binary mRS.
}

\subsection{Experimental setup} \label{sec:experimentalsetup}

\paragraph{Models}
We compare models with varying degrees of interpretability and flexibility 
for ordinal mRS prediction (see Tab.~\ref{tab:mods}). The goal is to obtain a model 
which achieves the highest possible prediction power while being adequately 
interpretable. In all models we choose $\pZ(\rz) = \expit(\rz)$, such that shift parameters 
in $\h$ can be interpreted as log odds-ratios. By comparing models 
based on tabular data, image data and a combination of both, we assess if tabular 
and image data carry complementary information and which of the two contains more information
for outcome prediction. As a baseline benchmark, we consider performance metrics of an 
unconditional model, which takes no input data and hence consists of a simple intercept 
(SI) only. This model predicts the prevalence of each outcome class. 
\revision{%
To assess binary mRS prediction, we consider the outcome as censored and sum up the predicted
probabilities of the respective ordinal model. The probability for a favorable outcome is the 
sum across the probabilities for classes 0 to 2, the probability for unfavorable outcome is the
sum across the probabilities for classes 3 to 6.
}

\begin{table}[!ht]
\centering
\caption{Summary of all models used for binary and ordinal functional outcome prediction in
the stroke data. If applicable, the transformation function is given. The model names
are combinations of the components (simple/complex intercept/shift), the subscript indicates
which modality enters which component, where \eg \lsx{} indicates that the tabular data is the
input. 
\revision{
SI: Simple intercept. CI: Complex intercept. LS: Linear shift. CS: Complex shift.
Note that \lsx{} includes all predictors, including pre-stroke mRS, whereas \lsmRS{}
contains pre-stroke mRS only.
}
} \label{tab:mods}
\resizebox{0.85\textwidth}{!}{%
\begin{tabular}{@{}llll@{}}
\toprule
\textbf{Outcome} & \textbf{Input data} & \textbf{Model name} & \textbf{Transformation function} \\ \midrule
Binary mRS & Images only & CI$_\B$-Binary & $\theta(\B)$ \\
\midrule
\multirow{8}{*}{Ordinal mRS} & None & SI & $\theta_k$ \\
& Tabular only & SI-LS$_{\rx}$ & $\theta_k - \rx^\top\shiftparm$     \\
& Tabular only & SI-CS\textsubscript{age}-LS$_{\tilde\rx}$ & 
    $\theta_k - \gamma(\rx_{\text{age}}) - \rx_{-\text{age}}^\top\shiftparm$     \\
& Images only & SI-CS$_\B$ & $\theta_k - \eta(\B)$ \\
& Images + tabular & SI-CS$_\B$-LS$_{\rx}$ & $\theta_k - \eta(\B) - \rx^\top\shiftparm$ \\ 
& Images + pre-stroke mRS & CI$_\B$-LS$_{\text{mRS}}$ & $\theta_k(\B) - \rx_{\text{mRS}}^\top\shiftparm$ \\
& Images + tabular & CI$_\B$-LS$_{\rx}$ & $\theta_k(\B) - \rx^\top\shiftparm$ \\
\cmidrule{2-4}
& Tabular only & GAM & $\theta_k - \gamma(\text{age}) - \rx_{-\text{age}}^\top\shiftparm$     \\
& Images only & CI$_\B$ & \\
\bottomrule
\end{tabular}}
\end{table}

We define an image-only model which is fitted using the binary mRS (CI$_\B$-Binary 
in Tab.~\ref{tab:mods}). This dichotomized version of the mRS can be viewed as a censored 
version of the ordinal mRS and can therefore be directly compared to all models fitted on 
the ordinal scale (see Tab.~\ref{tab:mods}). We fit the CI$_\B$-Binary model primarily
as a benchmark for the performance of the models that are trained for the ordinal but evaluated 
for the binary mRS.

The most interpretable model for the ordinal mRS is a linear
proportional odds model based on all tabular features. It consists of a
simple intercept and a linear shift in $\rx$ (SI-LS$_{\rx}$). 
The SI-CS$_{\code{age}}$-LS$_{\tilde\rx}$ model allows the outcome 
to depend on age in a non-linear way by estimating a potentially complex and continuous 
log odds-ratio function $x_{\text{age}}$. We additionally fit models depending 
on image data only (SI-CS$_\B$, CI$_\B$) and on a combination of image and tabular data 
(SI-CS$_{\B}$-LS$_{\rx}$ and CI$_{\B}$-LS$_{\rx}$ models). Integrating the images as 
complex intercept (CI$_{\B}$) rather than as complex shift term (CS$_{\B}$) allows
to increase model complexity further. In the image model CI$_\B$-LS$_{\text{mRS}}$, we 
additionally adjust for the pre-stroke mRS to achieve a fairer comparison between 
image-data-only and tabular-data-only.

\revision{%
\paragraph{Implementation} 
Simple intercept and linear shift terms for tabular features are modelled with fully
connected NNs without hidden layers. A fully connected NN with multiple hidden layers
is used to integrate age as complex shift term. The complex intercept and complex shift
terms for the images are modelled with a 3D CNN. In all models, the number of output
nodes is equal to six (since the mRS has seven levels) in NNs for intercept terms and
equal to one in NNs for shift terms. The last layer activation is always linear and no 
bias terms are used.

All models are trained by minimizing the negative log-likelihood (see Eq.~\ref{eq:nll}) 
using the Adam optimizer \citep{Kingma2015adam} with a learning rate of $5\times10^{-5}$ and a 
batch size of six. Augmentation of the image data is used to prevent overfitting. In addition, we 
use early stopping, \ie we select the model weights from the epoch which shows the smallest 
NLL on the validation data. More details on NN architectures, hyperparameters, augmentation 
procedure and software is given in Appendix~\ref{app:computation}.

\paragraph{Training and evaluation}
We randomly split the data six times into a train (80\%), validation (10\%) and test set (10\%).
This results in six fits for each model type (see table~\ref{tab:mods}) which allows us to 
assess the variation of the achieved test performance 
(see for example Fig.~\ref{fig:censored:mrs} A). For all models, that include the image data
as input, we perform transformation ensembling \citep{kook2022interpretable}.
For that, we train five models on the same data in each split. CNNs controlling the 
image term in the model are initialized randomly. 
Additional SI and LS$_\rx$ terms are initialized with the corresponding parameters of the  
SI-LS$_\rx$ model fitted on the same split. This results in an ensemble model (constructed 
from 5 members) in each of the six splits for each model type. 
}

\paragraph{Performance Measures} 
All models are mainly evaluated with proper
scoring rules \citep{gneiting2007strictly}. The score we consider primarily for
model comparison is the test negative log-likelihood \citep[NLL,][a.k.a.
log-score]{good1952rational}. We further assess the Brier score for the binary
outcome.
\revision{%
For the ordinal functional outcome, we calculate the ranked probability score as
an additional proper score \citep{brocker2007scoring}.
As measures of discriminatory ability, we compute AUC and accuracy for binary 
outcomes and quadratic weighted Cohen's $\kappa$ for the ordinal outcome
\citep{steyerberg2019}.
We construct 95\% bootstrap confidence intervals by taking $B = 1'000$ bootstrap
samples of size $n_{\rm test}$ of test predictions (\eg NLL contributions) for
each of the $S = 6$ random splits of the data, by computing the 2.5th, 50th, and
97.5th percentile of the $B$ bootstrap metrics averaged over the $S$ splits.
}

\section{Results and discussion} \label{sec:results}

We first present results for predicting and discriminating binary and ordinal mRS.
Then, we discuss how to interpret linear and non-linear model components.

\begin{figure}[!ht]
    \centering
    \includegraphics[width=0.95\textwidth]{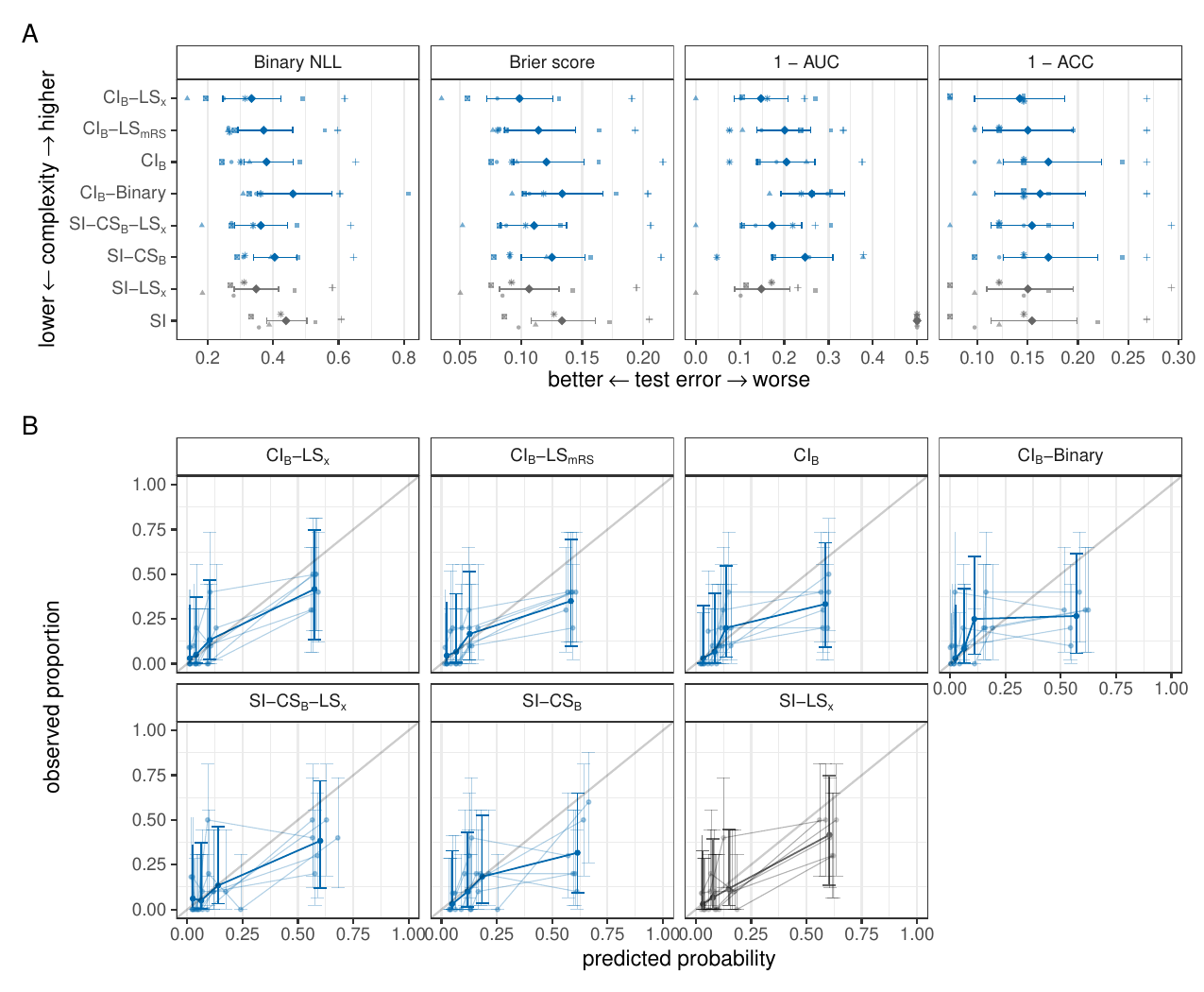}
    \caption{Test error (\textsf{A}) and calibration plots (\textsf{B}) 
    of transformation ensembles (blue) and reference models (grey) evaluated
    for the binary mRS outcome (mRS 0--2 \textit{vs.} mRS 3--6). In \textsf{A},
    the test performance 
    is quantified in terms of negative log-likelihood (NLL), Brier score,
    discrimination error ($1 - \operatorname{AUC}$) and classification error
    ($1 - \operatorname{ACC}$).
    From the test performance in the six random splits (indicated by different symbols) we 
    consider the
    average test error and 95\% bootstrap ($B = 1'000$) confidence intervals.
    \revision{%
    For the calibration plots in \textsf{B}, the predicted probabilities 
    are split at the 0.25, 0.5 and 0.75 empirical quantiles to produce the four bins
    for which the average predicted probabilities and the observed proportion of an
    unfavorable outcome are computed. The confidence interval is plotted at the midpoint
    of the respective bin.
    Average calibration across all six random splits are shown as thick line
    whereas the calibration of the single splits are shown as thin lines.
    }
    }
    \label{fig:censored:mrs}
\end{figure}

\paragraph{Binary mRS prediction}
The test performance and calibration plots of all models from Tab.~\ref{tab:mods} 
evaluated for the binary mRS are summarized in Fig.~\ref{fig:censored:mrs}.
We first compare models which only include the
image modality and only differ in the number of classes (CI$_\B$-Binary trained
with two \emph{vs.} CI$_\B$ trained with seven classes). The CI$_\B$-Binary model
shows a worse average performance and a higher variability in predictions across
the six random splits compared to the CI$_\B$. This highlights the importance for
training with all available class levels rather than with a dichotomized version
of the outcome -- whenever possible. The average performance of the CI$_\B$-Binary
is similar to that of the unconditional model (SI) indicating that the model has
primarily learned the class frequencies. Decreasing model complexity by modelling
the image data with a complex shift (SI-CS$_\B$) rather than with a complex
intercept (CI$_\B$) leads to a comparable performance. Both models, SI-CS$_\B$
and CI$_\B$, achieve a better average prediction performance than the unconditional
model (SI) indicating that the image data contains information for mRS binary
prediction.

The most interpretable model based on tabular features only (SI-LS$_\rx$) shows a
better prediction performance than all models based on image data only
(CI$_\B$, CI$_\B$-Binary, SI-CS$_\B$) in terms of NLL and Brier Score. Like the
models based on image data only, the SI-LS$_\rx$ outperforms the unconditional
model (SI, Fig.~\ref{fig:censored:mrs}). This indicates that not only image but also
tabular data is useful for binary mRS prediction. For a fairer comparison of
image-data-only \emph{vs.} tabular-data-only, we adjust for pre-stroke mRS in the
most flexible image model. In this comparison, the baseline adjusted model
(CI$_\B$-LS\textsubscript{mRS}) shows a performance similar to the unadjusted model
(CI$_\B$).

The semi-structured models incorporating both image and tabular data (\ciblsx and 
\sicsblsx) achieve a similar or slightly better average performance than the model
including tabular data only (\silsx, see Fig.~\ref{fig:censored:mrs}).
\revision{%
\ciblsx does not assume proportional odds for the image modality and outperforms
\sicsblsx on some splits. The latter assumes proportional odds for both tabular 
and image data. Overall, there is no convincing evidence that combining tabular and 
imaging data in a \ciblsx model improves binary mRS prediction. The added image 
information increases variability in prediction performance.
}

Scores highlighting discriminatory ability of the models (AUC and accuracy) show
similar results. Slight differences in the ranking of models are possible because
these measures are improper scoring rules \citep{gneiting2007strictly}.
Note that SI has no discriminatory ability (AUC $=$ 0.5) because it always predicts 
the most frequent class (mRS 0). The relative test performance to the benchmark 
\silsx{} model (\ie the differences in performance within splits) can be found in
Appendix~\ref{subsec:rel}.

\revision{%
Well-calibrated predictions are hard to achieve for highly imbalanced outcomes.
The calibration plots in Fig.~\ref{fig:censored:mrs} show no substantial evidence for
miscalibration. However, all models seem to slightly over-predict the probability for 
an unfavorable outcome.
This effect is most pronounced in the models based on image data only (\cib, \cib-Binary). 
The semi-structured and tabular data-only models show a slightly better calibration.
}

\begin{figure}[!ht]
    \centering
    \includegraphics[width=0.7\textwidth]{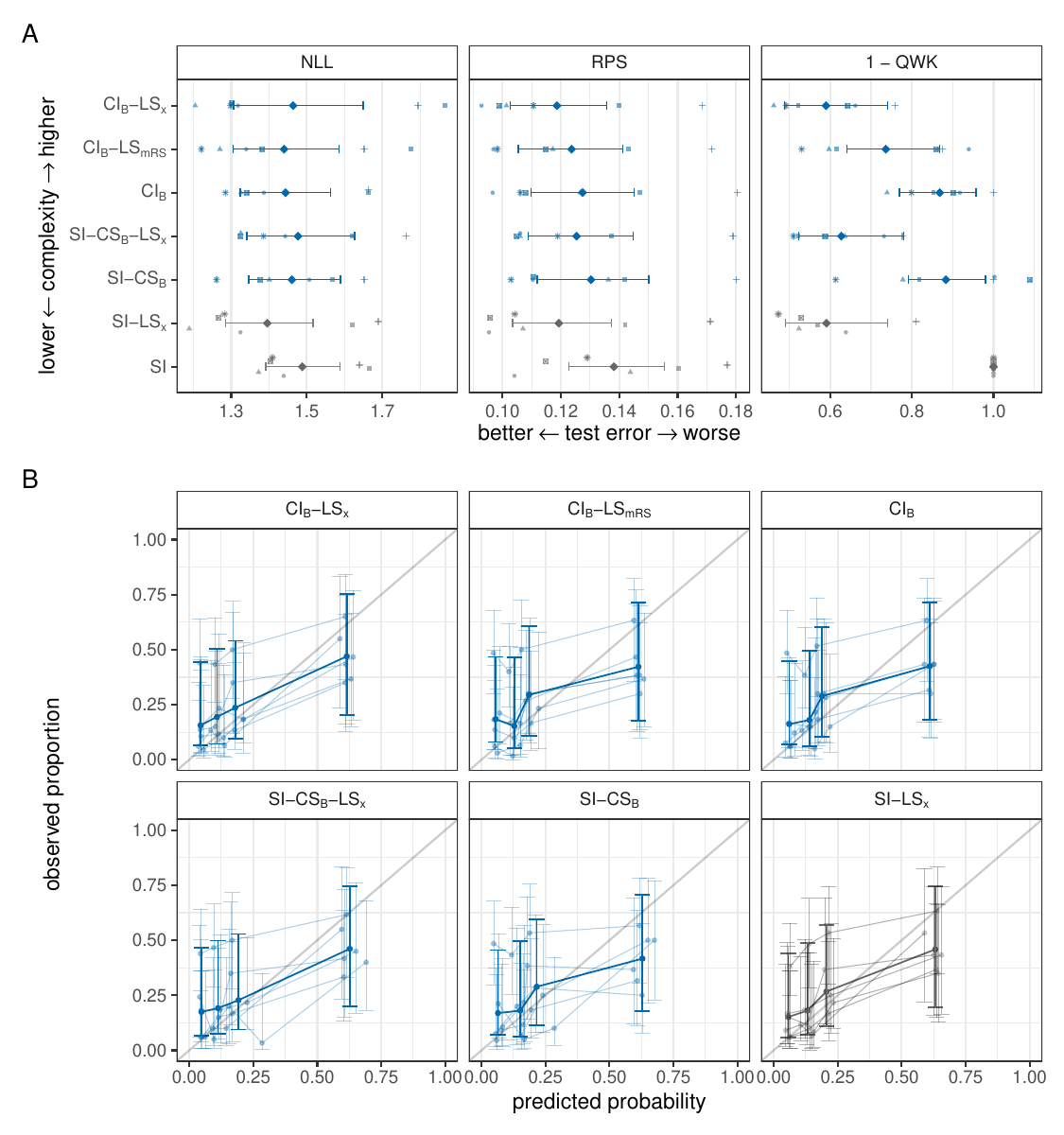}
    \caption{Test error (\textsf{A}) and calibration plots (\textsf{B}) 
    of transformation ensembles (blue) and reference models (grey)
    evaluated for the ordinal mRS outcome. In \textsf{A}, test error is quantified 
    in terms of negative log-likelihood (NLL), ranked probability score (RPS),
    and discrimination error ($1-\operatorname{QWK}$). The average test error 
    and 95\% bootstrap ($B = 1'000$) confidence intervals are depicted for six 
    random splits of the data (indicated by the different symbols). 
    \revision{%
    For the calibration plots in \textsf{B}, the predicted probabilities 
    are split at the 0.25, 0.5 and 0.75 empirical
    quantiles to produce the four bins for which the average predicted probabilities 
    and the observed proportion of an unfavorable outcome are computed.
    The confidence interval is plotted at the midpoint of the respective bin. 
    Average calibration across all six random splits is shown as thick line whereas 
    the calibration of the single splits is shown as thin lines. 95\% confidence 
    intervals are averaged across classes and splits.
    }
    }
    \label{fig:ordinal:mrs}
\end{figure}

\paragraph{Ordinal mRS prediction}
Fig.~\ref{fig:ordinal:mrs} summarizes the test performance and calibration plots
for all models in Tab.~\ref{tab:mods} trained and evaluated for the ordinal mRS.
As in the binary case, the models based on image (CI$_\B$, SI-CS$_\B$) and 
tabular data only (SI-LS$_\rx$) show better average prediction performances in 
terms of NLL, RPS and QWK than the unconditional model (SI). And again, the most
interpretable model based on tabular data only (SI-LS$_\rx$) outperforms the more
flexible black box image-only models, indicating that tabular features contain
more information for ordinal mRS prediction than the images (at the available 
sample with only 407 patients). As in the binary case,
we find no substantial evidence that using tabular and image data together in a 
semi-structured model (CI$_\B$-LS$_\rx$, CI$_\B$-LS\textsubscript{mRS} or 
SI-CS$_\B$-LS$_\rx$) improves average test performance compared to SI-LS$_\rx$ 
(see Fig.~\ref{fig:ord:rel}). 

In terms of calibration (Fig.~\ref{fig:ordinal:mrs}\textsf{B}) we again observe 
that all models over-predicted the probability for an unfavorable outcome.

Overall, we can not draw a definitive conclusion about which data modality 
(tabular or image data) is more useful for functional outcome prediction and 
if adding image to tabular data aids mRS prediction. The confidence intervals 
overlap largely and average test performance is similar. In particular, this
can be attributed to the small sample size. In Appendix~\ref{app:sample_size}, 
we conclude that collecting more data could further enhance performance.
When we artificially reduce sample size via sub-sampling and refit all 
models, we find no evidence of plateauing prediction performance. However,
no differential increase in prediction performance is observed for the
tabular-data-only model compared to the most complex DTM.

\begin{figure}[!ht]
    \centering
    \includegraphics[width=0.8\textwidth]{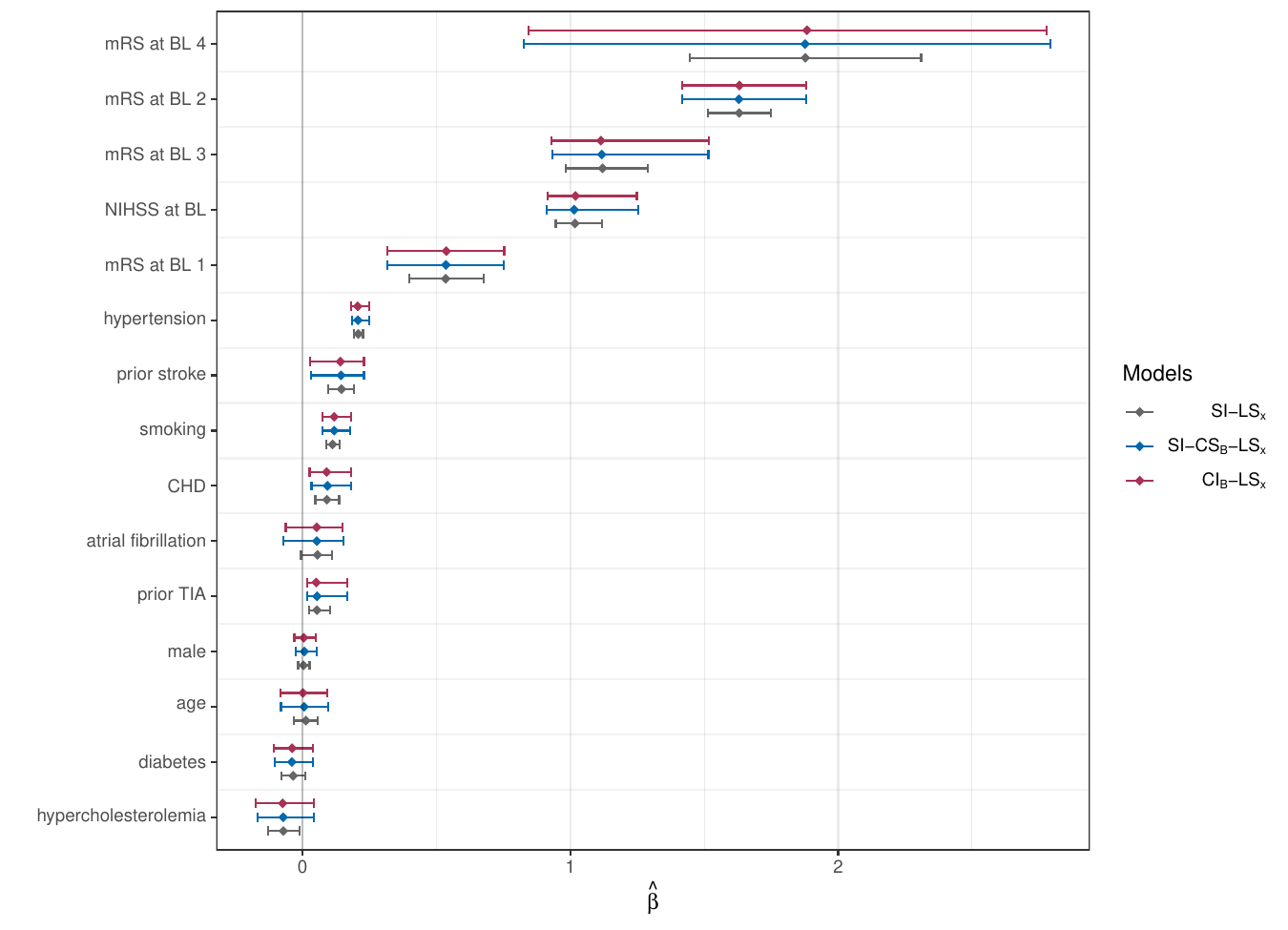}
    \caption{Pooled log odds-ratios ($\hat\betavec$) across all six random 
    splits and 95\% bootstrap ($B = 1'000$) confidence intervals for all models
    with linearly included tabular features (see Section~\ref{sec:data}).
    With the exception of age and NIHSS, all features are categorical and the plot 
    shows log odds-ratios with respect to the reference level (note that 
    the largest observed pre-stroke mRS is 4). The coefficients are based
    on standardized features and sorted with respect to an increasing order.
    }
    \label{fig:ordinal:effect_sizes}
\end{figure}

\paragraph{Interpretation of model parameters}
Fig.~\ref{fig:ordinal:effect_sizes} visualizes the effect sizes of the tabular 
features in the linear shift terms LS$_\rx$ of different models. 
Because the logistic distribution is chosen for $\rZ$, the coefficients 
$\betavec$ in the linear shift term are interpretable as log odds-ratios. 
Comparing tabular-data-only models with semi-structured models shows that
adjusting for the images (in CI$_\B$-LS$_\rx$ or SI-CS$_\B$-LS$_\rx$) 
changes the $\hat\betavec$ estimates only slightly.
The log odds-ratios are comparable across all variables because the
variables are standardized. Thus, the effect sizes reflect a change
in log-odds of a worse outcome for a one standard deviation increase in
the respective variable. Accordingly, Fig.~\ref{fig:ordinal:effect_sizes}
shows that the strongest prognostic factors are the pre-stroke mRS and NIHSS 
on admission.
This is expected when predicting three months mRS. The pre-stroke mRS captures functional disability
of a patient before stroke while NIHSS measures stroke severity on
admission. This additionally emphasizes the importance for being able to adjust for
pre-stroke mRS.

\revision{%
Similar to both linear and complex shift terms, complex intercepts of categorical 
predictors are directly interpretable. Here, cumulative baseline log-odds of
the outcome are estimated for each stratum of the predictor. Thus, differences
in the complex intercepts can be interpreted as class-specific log-odds ratios
\citep{Buri2020}.
For continuous predictors or images, this simple interpretation is lost to an 
extent which depends on the complexity of the neural network component that is 
modelling the complex intercept term.
}

\revision{%
Alongside interpretation, quantifying uncertainty in both predictions and
parameters is of high importance, but generally difficult to achieve in deep
learning models \citep{wilson2020bayesian}. The use of transformation ensembles 
and random splits allows uncertainty quantification for the coefficient estimates 
in terms of bootstrap confidence intervals. This way, both aleatoric and epistemic
uncertainty are captured. Note, that the model coefficients of the five (members) 
times six (splits) are repeatedly sampled and that the models are not additionally 
refitted to obtain the 95\% confidence intervals.}

\begin{figure}[!ht]
    \centering
    \includegraphics[width=0.65\textwidth]{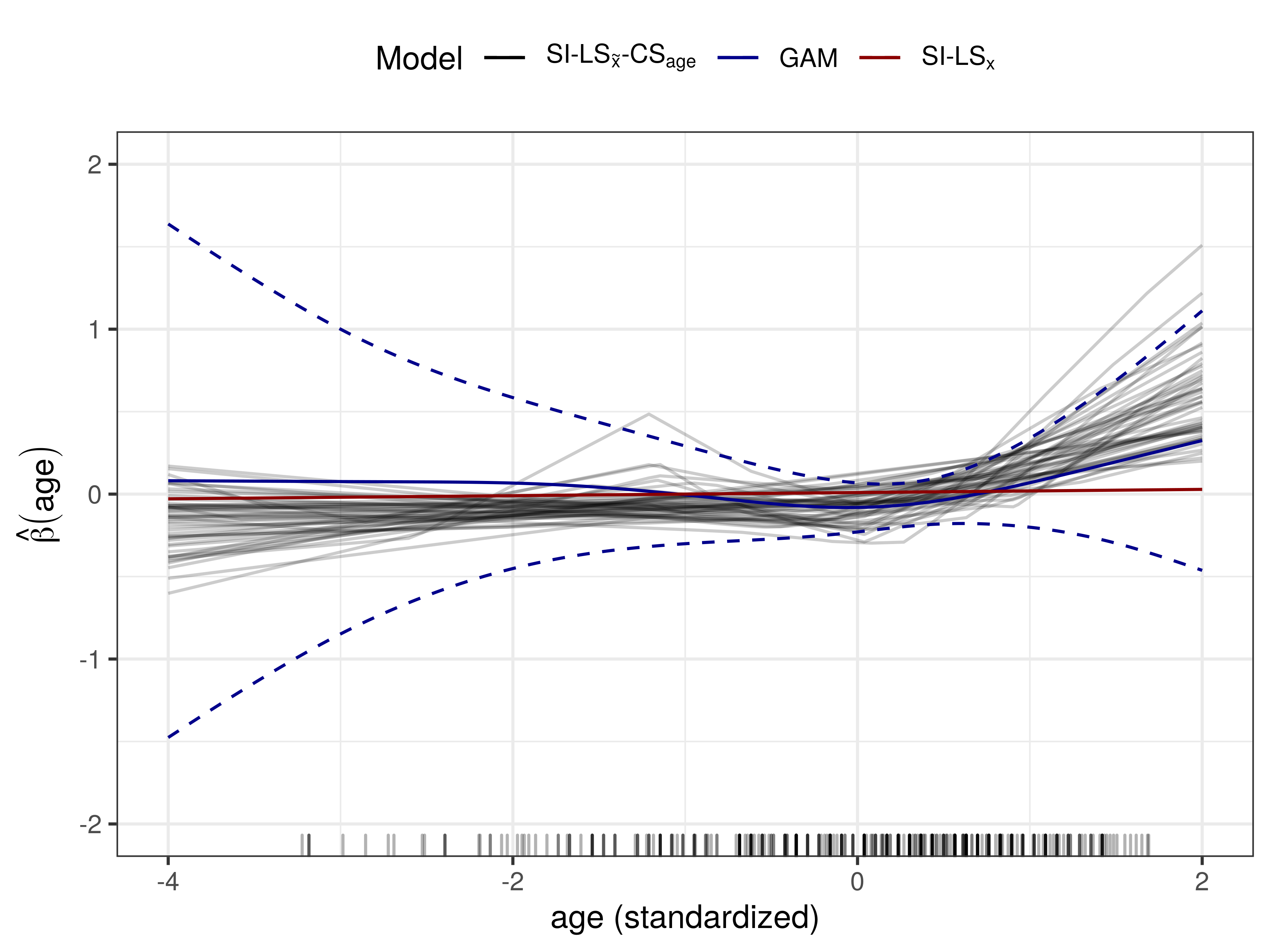}
    \caption{The smooth log-odds function for age fitted by a GAM (a generalized 
    additive proportional odds model using the \pkg{mgcv} package), depicted as 
    blue solid line along with point-wise 95\% confidence band (dashed lines), and
    by a DTM (SI-LS$_{\tilde{\rx}}$-CS\textsubscript{age}) fitted on 50 bootstrap 
    samples, depicted as grey lines. In addition,
    the linear effect of a SI-LS$_\rx$ is displayed (red line). Although both models,
    GAM and DTM, allow for a non-linear effect in age, there is no evidence against 
    a linear effect of age.}
    \label{fig:ordinal:mrs:gam}
\end{figure}
To investigate if assuming a linear age effect is appropriate, we evaluate
models including the age effect with a flexible function, $\theta_k-\gamma(x_{\text{age}})-
\rx_{-\text{age}}^\top\shiftparm$. We show the results of a GAM 
(a generalized additive proportional odds model) and a DTM 
(SI-LS$_{\rx}$-CS\textsubscript{age}) which depict the estimated age effect
function as shown in Fig.~\ref{fig:ordinal:mrs:gam}. 
The GAM and the DTM agree in the functional form of the effect, which 
is constant up to a standardized age of 0.5 (corresponding to an age of 75 years) and 
then increases the odds for a worse outcome. However, there is no evidence against
a linear effect when we consider the point-wise confidence band for the GAM.
Note how the GAM enforces smoothness of the estimated function, whereas the
neural network produces a piece-wise linear estimate.

\section{Summary and outlook} \label{sec:outlook}

DTMs provide a novel and flexible way to integrate multi-modal data for interpretable 
prediction models for various kinds of outcomes. We demonstrate the potential of 
DTMs on a semi-structured data set with an ordinal outcome (mRS) describing the 
functional disability of stroke patients three months after hospital admission.
We discuss how the best trade-off between interpretability and flexibility can be 
achieved. In essence, we follow the top-down model approach to model building for 
TMs \citep{hothorn2018topdown}. 
\revision{%
By investigating the interpretable model parameters, we judge the relative importance
of the predictors and show that in a baseline-adjusted DTMs, the base-line mRS is the 
variable with the most relevant predictive effect. We also investigate the question, 
which input modality is most important for functional outcome prediction and whether 
predictive performance in terms of NLL and calibration can be improved by including
both tabular and imaging data. While for binary mRS prediction, models seemed to 
slightly benefit from the addition of brain imaging data, this is not observed for 
ordinal mRS prediction. In general, a definitive judgement on whether the images contain 
information to aid mRS prediction cannot be made. This is because, all results have 
to be interpreted conditional on (i) the small sample size and (ii) limited computation 
time for joint hyper-parameter tuning.
}
When artificially increasing the sample size up to the available 407 patients, there
is no evidence for differential performance gain of the semi-structured over 
tabular-data-only models. However, extrapolating these results to larger sample sizes
is in general extremely difficult.

\revision{%
In general, deep neural networks (including DTMs) are difficult to train with
limited sample size and require a carefully chosen optimization procedure. For
instance, transfer learning in terms of adapting the weights of a CNN that is 
already trained on a different data set by re-training it with the data of 
interest potentially improves predictive performance even with smaller sample sizes.
However, methods for transferring the weights of well-known 2D CNN architectures 
to their 3D counterparts did not improve predictive performance in our application
(results not shown). In general, it is difficult to access weights of trained 3D CNNs 
to then fine-tune the models.
}

For ordinal functional outcome prediction in our cohort, the model SI-LS$_\rx$ 
seemed to be most appropriate when including tabular features only and modelling them 
as linear effects. Here, classical statistical inference provides uncertainty measures 
(confidence intervals) and the model is fully interpretable. Using semi-structured 
models, including tabular and brain imaging data, improved binary mRS prediction to some 
extent. However, including images as a complex intercept or complex shift reduced 
interpretability of the model and increased variability.

TMs also work naturally for other kinds of outcomes, such as survival times,
which often feature censored observations \citep[\eg][]{hothorn2018most}. Because 
the dichotomized mRS could be viewed as a censored version of the ordinal mRS, the 
very same models, trained on ordinal outcomes, can also be used for different 
dichotimizations (or binnings) of the ordinal outcome, without the need to re-fit 
the models on the binned outcome \citep{lohse2017continuous}.

In summary, being able to fit distributional regression models with complex outcome 
types and multi-modal input data and following statistical principles for model 
building opens up vast areas of applications. Especially in medicine, these models 
have the potential to aid decision making, because of their state-of-the-art 
prediction performance and transparency.

\section*{Acknowledgements}
The research of LH, LK, SW and BS was supported by Novartis Research Foundation 
(FreeNovation~2019) and by the Swiss National Science Foundation 
(grant no. S-86013-01-01 and S-42344-04-01).
OD was supported by the Federal Ministry of Education and Research of Germany 
(grant no. 01IS19083A).


\vskip 0.2in
\bibliographystyle{plainnat}
\bibliography{bibliography} 


\appendix
\renewcommand{\thesection}{\Alph{section}}
\counterwithin{figure}{section}
\renewcommand\thefigure{\thesection\arabic{figure}}
\counterwithin{table}{section}
\renewcommand\thetable{\thesection\arabic{table}}

\section{Computational details} \label{app:computation}

For reproducibility, all code is made publicly available on GitHub
\url{https://www.github.com/LucasKookUZH/dtm-usz-stroke}.

\paragraph{Neural Network architectures}
The simple intercept terms are modelled with a fully connected single-layer NN without 
hidden layers and linear activation. No bias term is used. The number of output nodes
is always equal to the number of classes minus one while the input is a vector of ones. 

The linear shift terms for the tabular data are modelled with fully connected NNs 
without hidden layers and a linear function as activation. No bias term is used. 
The number of input units is equal to the number of tabular features while the number
of units in the last layer is equal to one.

The complex shift term for the variable age is modelled with a fully connected NN 
with two hidden layers with 16 units each, ReLU activation function and $L_2$ 
regularization. The number of units in the last layer is equal to one and the activation 
function in this layer is linear. 

The complex intercept and complex shift terms for the image data are modelled with 
a 3D CNN. The convolutional part of the 3D CNN consists of four convolutional blocks 
including a convolutional layer with filter size $3\times 3\times 3$ and a max pooling 
layer of size $2 \times 2 \times 2$. The first two layers use 32 filters, the following 
two use 64 filters. The subsequent fully connected part consists of two fully connected 
layers with 128 filters each, that are separated by a dropout layer with dropout rate 0.3.
The activation function in all layers, expect the last one, is the ReLU non-linearity. 
In case the image data is included as complex intercept term, the number of units in the last 
layer is equal to the number of classes minus one, \ie one when we predict the binary mRS 
and 6 when we predict the ordinal mRS. When integrated as complex shift term, the number 
of units in the last layer is equal to one. The activation function in the last layer 
is linear.

\paragraph{Training}
All models are fitted with stochastic gradient descent using the Adam optimizer 
\citep{Kingma2015adam} with a learning rate of $5\times10^{-5}$ and a batch size of six. 
We then use the model with the best performance on the validation data in terms of NLL.
For all experiments, the 3D images were augmented in x- and y-direction prior to every 
epoch using the parameters in Tab.~\ref{tab:aug}.

\begin{table}[!ht]
    \centering
    \caption{Parameter values for augmentation.}
    \label{tab:aug}
    \begin{tabular}{lr}
    \toprule
    \bf Parameter & \bf Value \\
    \midrule
         rotation range & 20 \\
         width shift range & 0.2 \\
         height shift range & 0.2 \\
         shear range & 0.15 \\
         zoom range & 0.15 \\
         fill & nearest \\
    \bottomrule
    \end{tabular}
\end{table}

All models are implemented in \proglang{R}~4.1.2 \citep{pkg:base}. The models are 
written in \texttt{Keras} based on \texttt{TensorFlow} backend using \texttt{TensorFlow}
version 2.2.0 \citep{chollet2015keras,tensorflow2015} and trained on a GPU.
Linear proportional odds models and generalized additive proportional odds
models are fitted using \texttt{tram::Polr}
\citep{pkg:tram} and \texttt{mgcv::gam} \citep{pkg:mgcv}, respectively. 

\revision{
\section{Additional results}

Here, we present descriptive statistics and additional results.

\subsection{Baseline characteristics}

Fig.~\ref{fig:predictors} shows the distribution of predictors stratified
by the outcome (90 day mRS) in the stroke data set.

\begin{figure}[!ht]
    \centering
    \includegraphics[width=0.95\textwidth]{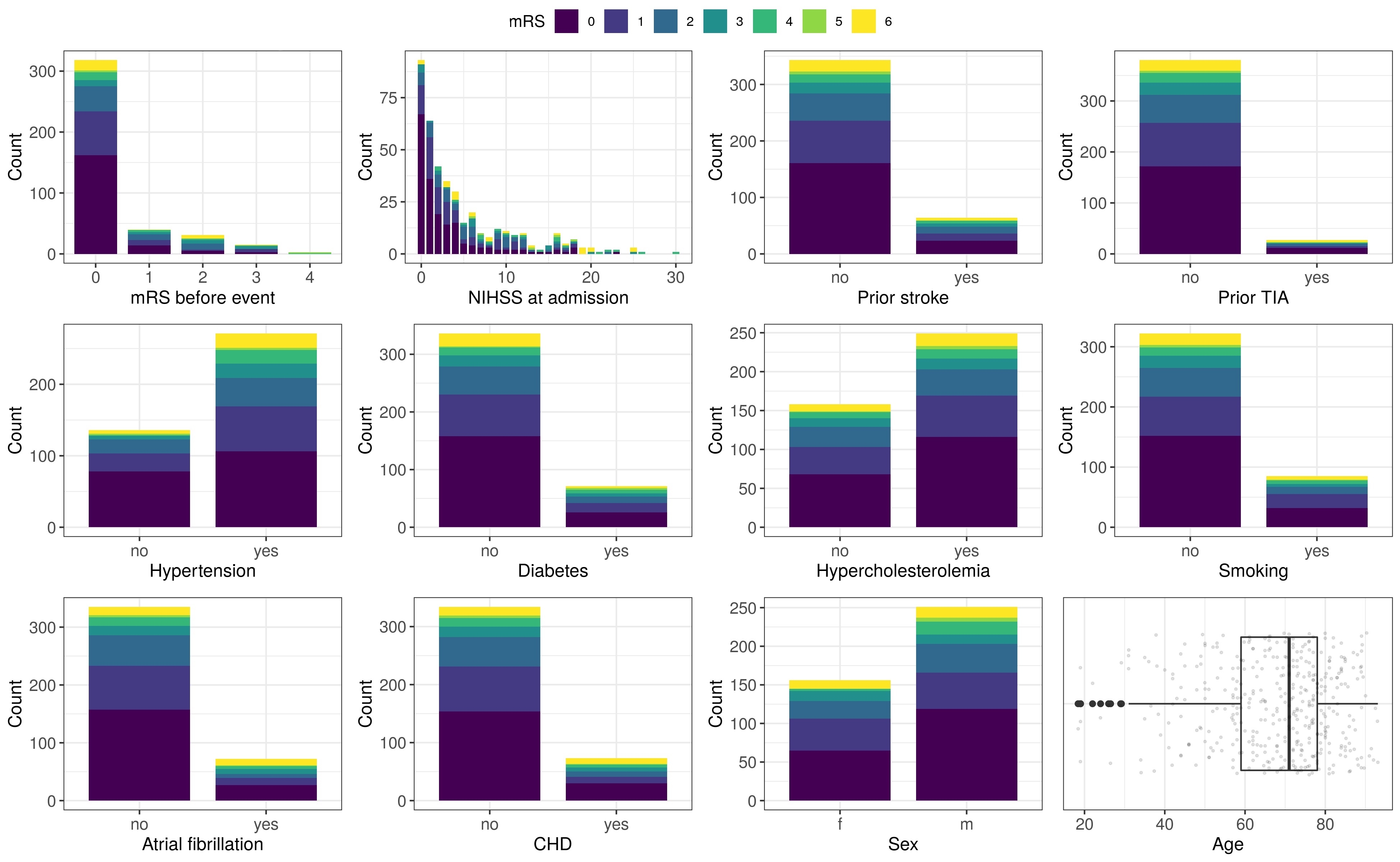}
    \caption{\revision{Baseline characteristics stratified by 90 day mRS
    in the stroke data set.}}
    \label{fig:predictors}
\end{figure}

\subsection{Test errors relative to reference model} \label{subsec:rel}

Figg.~\ref{fig:bin:rel} and \ref{fig:ord:rel} show the test errors
relative to the reference \silsx{} model evaluated on the
binary and ordinal mRS, respectively. After removing the
between-split variance, none of the semi-structured models
improve significantly upon the performance of the \silsx{} model.
Since the \silsx{} performance was not bootstrapped (the constant 
split-wise mean was subtracted within split) there is no variance in the 
average AUC and QWK (because the \si model does not have any discriminatory
ability, \ie AUC~$= 0.5$ and QWK~$=0$) across splits for the unconditional 
SI model.

\begin{figure}[!ht]
    \centering
    \includegraphics[width=0.95\textwidth]{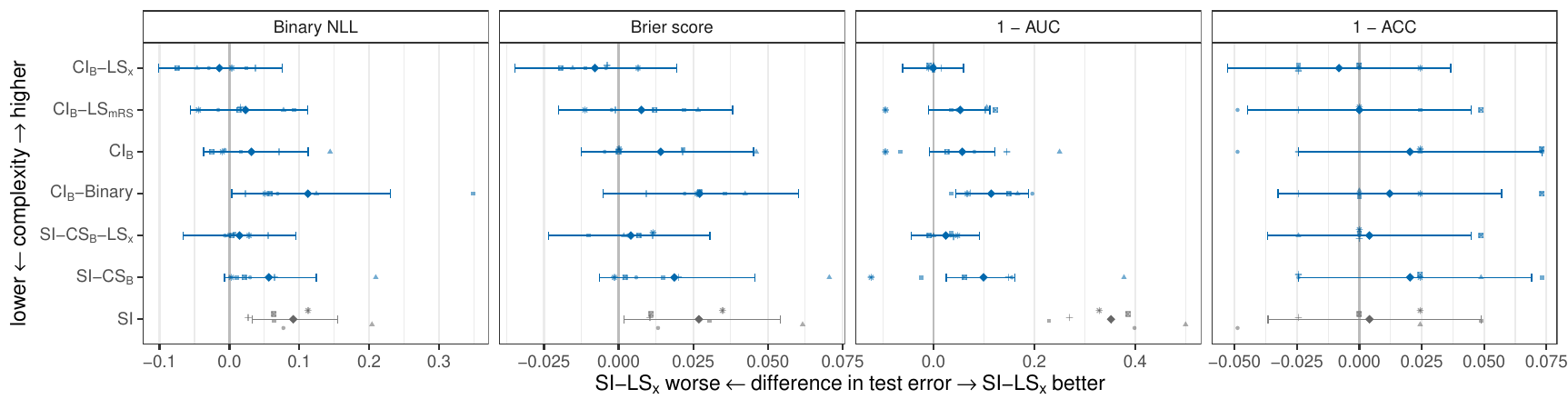}
    \caption{\revision{Test error of transformation ensemble models (blue) and 
    reference model (grey) evaluated for the binary mRS outcome 
    (mRS 0--2 \textit{vs.} mRS 3--6) relative to the test error of the
    benchmark \silsx{} model (\ie a difference of 0 indicates the same
    performance as the \silsx{} model).
    The test error is quantified in terms of binary negative
    log-likelihood (NLL), Brier score, $1-$ area under the ROC curve 
    (AUC) and classification error ($1-\operatorname{ACC}$).
    The average test error and 95\% bootstrap ($B = 1'000$) confidence
    intervals (CI) are depicted for six random splits 
    (indicated by different symbols). The CIs are calculated by substracting 
    the fixed \silsx{} performance per split.}}
    \label{fig:bin:rel}
\end{figure}

\begin{figure}[!ht]
    \centering
    \includegraphics[width=0.7\textwidth]{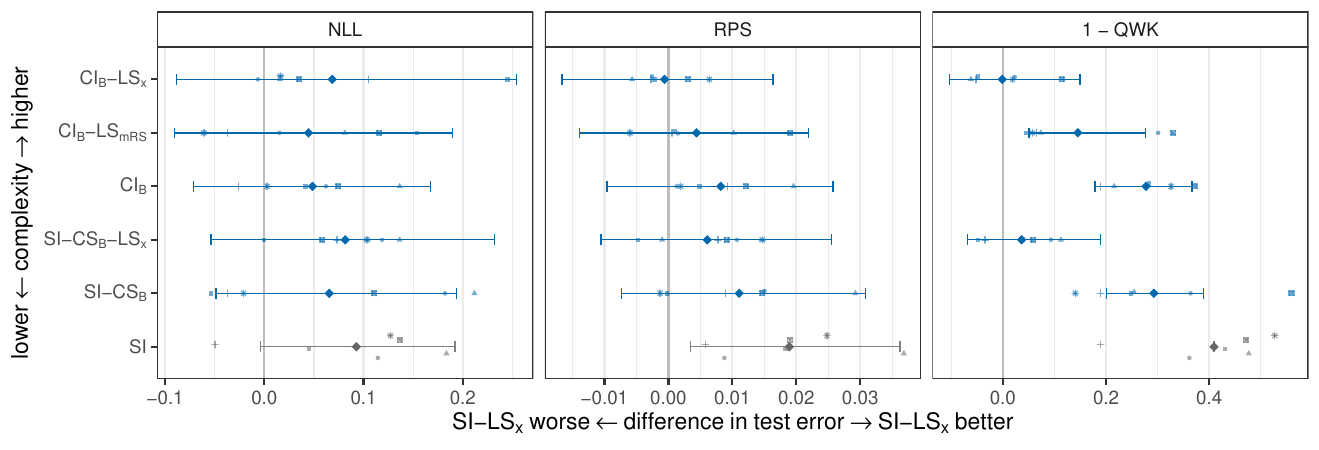}
    \caption{\revision{Test error of transformation ensemble models (blue) and 
    reference model (grey) evaluated for the ordinal mRS outcome 
    relative to the test error of the benchmark \silsx{} model 
    (\ie a difference of 0 indicates the same performance as the 
    \silsx{} model).
    The test error is quantified in terms of negative
    log-likelihood (NLL), ranked probability score (RPS)
    and discrimination error ($1-\operatorname{QWK}$).
    The average test error and 95\% bootstrap ($B = 1'000$) confidence
    intervals (CI) are depicted for six random splits 
    (indicated by different symbols). The CIs are calculated by substracting 
    the fixed \silsx{} performance per split.}}
    \label{fig:ord:rel}
\end{figure}
}

\subsection{Sample size}\label{app:sample_size}

\begin{figure}[!ht]
    \centering
    \includegraphics[width=0.95\textwidth]{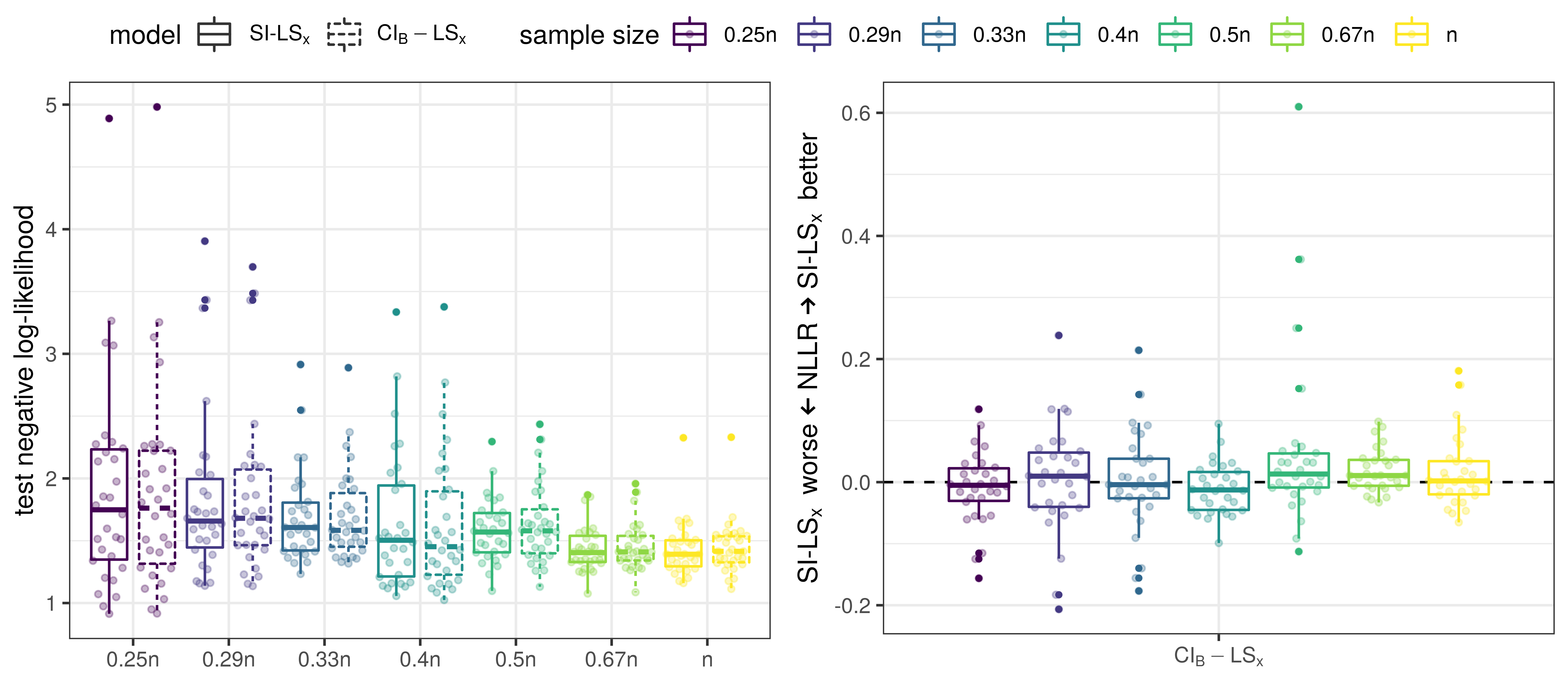}
    \caption{Test performance versus sample size achieved by a subsampling experiment. 
    The semi-structured CI$_\B$-LS$_\rx$ model and the proportional odds model 
    SI-LS$_\rx$ are compared. Both models are fitted to 30 random sub-samples of 
    seven different sample sizes of the original sample size ($n = 407$) and the 
    test NLL is recorded. Both models benefit from increasing sample size.
    The right panel displays the differences in NLL within split for a given sample 
    size (\ie the negative log-likelihood ratio NLLR).}
    \label{fig:ordinal:mrs:samplesize}
\end{figure}

Deep learning typically requires thousands of training images to excel at
prediction performance over conventional statistical models \citep{goodfellow2016deep}.
However, our cohort, like most medical data sets, contained 
much fewer observations ($n = 407$). To study if collecting more data was a
promising approach to enhance the model performance, we artificially reduced sample 
size by sub-sampling and refitted the models (see Fig.~\ref{fig:ordinal:mrs:samplesize}). 
In this experiment, the
sample size is artificially reduced via sub-sampling of varying sizes and then the
prediction performance was measured on a hold-out set of the reduced data set.
Sub-sampling is repeated for seven sample sizes and then 30
train/validation/test splits (with a ratio of 8:1:1) are used for fitting and evaluating the
semi-structured CI$_\B$-LS$_\rx$ and tabular-data-only model SI-LS$_\rx$. 
We observe that the prediction performance, \ie the test NLL, improves for both models 
with increasing sample size, indicating that the performance may further increase with 
increasing sample size (left panel of Fig.~\ref{fig:ordinal:mrs:samplesize}). 
Directly comparing the performance of both models for the individual 
splits suggests no evidence that adding the image information to the model that 
contained the tabular data as input improves prediction performance. 
The negative log-likelihood ratio fluctuates around zero and no trend with 
increasing sample size is observable (right panel in Fig.~\ref{fig:ordinal:mrs:samplesize}).

\end{document}